\documentclass[twocolumn]{aastex631}

\usepackage{amssymb}

\begin{document}

\title{The hot accretion flow evolution in the black hole X-ray binary MAXI J1348-630 }

\author[0000-0003-3003-866X]{Hanji Wu}
\affiliation{Department of Astronomy, School of Physics and Technology, Wuhan University, Wuhan 430072, China}

\author[0000-0003-3901-8403]{Wei Wang}
\affiliation{Department of Astronomy, School of Physics and Technology, Wuhan University, Wuhan 430072, China}
\correspondingauthor{Wei Wang}
\email{wangwei2017@whu.edu.cn}


\begin{abstract}

MAXI J1348-630 as a low-mass black hole binary system in the Galaxy showed an X-ray outburst in 2019. We analyzed the Insight-HXMT spectral data in the low hard state (LHS) and intermediate state (IS) during the outburst from MJD 58510 to 58519 at the energy band from 2 keV to 100 keV. During the entire process, a thin disk extending to the innermost stable circular orbit (ISCO) from a large truncated disk (truncated radius $> 5$ ISCO) suggests the corona geometry evolution. There exist time lags between radio and hard X-ray flux peaks: the 30 - 100 keV flux about 5 days ahead of radio flux, 11-30 keV flux about 4 days ahead, and reflection fraction about 2 days ahead, the accretion disk approaching the ISCO about 1 day before radio peak. This disk-corona-jet coupling and evolution suggest the corona containing two phases of cold dense material and hot gas, with high temperature region of corona cooling fast. The strong radio emission accompanying with a thin accretion disk of a relatively high accretion rate favors magnetic tower jet mechanism.

\end{abstract}


\section{Introduction}

Low-mass black hole X-ray binaries (LMBHXB) consist of a black hole (BH) accreting material from its low-mass comparison star through Roche lobe overflow \citep{savonije1978roche,abramowicz1988slim,lovelace1999rossby}. The primary radiation of this system can be characterized by multi-temperature black body spectra in soft X-ray range within 7 keV \citep{shakura1973black,novikov1973astrophysics}, while at the hard X-ray range, the spectra can be modeled by a cutoff power-law component that is attributed to the high-temperature plasma (up to $\sim 100\ \rm keV$, named as corona based on \citealt{sunyaev1980comptonization,sunyaev1985comptonization}). With the high-resolution spectra, the cutoff power-law component can't model the hump over 20 keV, and the iron emission lines, which can be well explained by the Comptonized photons reflected from the accretion disk \citep{fabian2016innermost}. Besides, the relativistic effects would make the iron emission line broad and asymmetric \citep{fabian1989x,reynolds2003fluorescent}. These spectral components would help to understand the geometry and the motion of the disk-corona near the BH\citep{niedzwiecki2016lamppost}. 

At the duration of the LMBHXB outburst, the flux in X-ray bands could change over an order of magnitude, the shape of the hardness intensity diagram (HID) usually obeys the 'Q' feature evolving from the low hard state (LHS) to the high soft state (HSS), through the intermediate state (IS) divided into hard intermediate state (HIS) and soft intermediate state (SIS) \citep{mcclintock2006spin,belloni2000model,fender2004towards}. The intensity of the source rises sharply from the LHS to HIS and continues into HSS, then, the intensity falls back into LHS before vanishing into quiescence. With state change, the spectra component of the disk and the corona evolve in different patterns. The power-law component produced by the corona decreases from LHS to HSS and the multi-temperature black body component made by the disk increases from LHS to HSS \citep{done2007modelling}.

The Monitor of All-sky X-ray Image (MAXI) using the onboard Gas Slit Camera (GSC) \citep{matsuoka2009maxi} discovered the X-ray source MAXI J1348-630 on 2019 January 26 \citep{yatabe2019maxi}, which is classified as LMBHXB candidate by the spectral features, the mass of the central celestial body, and the intensity evolution \citep{carotenuto2019meerkat,denisenko2019optical,jana2019preliminary,kennea2019maxi,russell2019atca,sanna2019nicer,yatabe2019maxi,belloni2020time,tominaga2020discovery}. The MAXI J1348-630 shows the 'Q'-shape spectral feature during the main outburst in 2019 from the Swift/XRT data \citep{tominaga2020discovery}, the NICER data \citep{zhang2020nicer}, and the Insight-HXMT data \citep{2019ATel12470....1C}. \cite{lamer2021giant} discovered a giant dust scattering ring centered on MAXI J1348-630 based on the SRG/eROSITA image, and estimated the mass of black hole with $11\pm 2\rm M_{\odot}$ at the distance of $3.39\pm 0.34\rm kpc$ by joint analysis of the XMM-Newton, MAXI, Gaia data. The Swift and MAXI data during the 2019 main outburst was modeled by a two-component advective flow model estimating the black hole mass to be $9.1^{+1.6}_{-1.2}\rm M_{\odot}$ \citep{jana2020accretion}. The radio data from the Australian Square Kilometre Array Pathfinder (ASKAP) and MeerKAT suggested the distance at $2.2^{+0.5}_{-0.6}\rm kpc$ by modeling the H\uppercase\expandafter{\romannumeral1} absorption spectra \citep{chauhan2021measuring}. Radio observations give the eject speed $\ge 0.69\rm c$, and the open angle of the jet $\le 6^{\circ}$ \citep{carotenuto2021black}, furthermore, the modeling of this jet's synchrotron emission shows the initial Lorentz factor $\Gamma_{0}=1.85^{+0.15}_{-0.12}$ with the kinetic energy $E_{0}=4.6^{+20.0}_{-3.4}\times 10^{46}\rm erg$ which is greater than most LMBHXBs\citep{carotenuto2022black}. In hard state and hard intermediate states, the magnetic launching disk wind is discovered by an absorption line around $\sim 7\rm keV$ coming from highly ionized iron with a velocity of $\sim 10^{4}\rm km\ s^{-1}$ \citep{wu2022accretion}, and the disk wind of the velocity $\sim 10^{3}\rm km\ s^{-1}$ is also discovered in soft state with optical and infrared energy band  \citep{panizo2022discovery}. 

The inclination angle of this source has also been studied widely. \cite{carotenuto2022modelling} estimated a $29^\circ \pm 3^\circ$ inclination by modeling the jet using the radio data, and \cite{carotenuto2021black} obtained an upper limit of a $46^\circ$ inclination from the radio image. The analysis of the reflection component in the X-ray spectra gave a range of inclination from $29^\circ$ to $40^\circ$ \citep{jia2022detailed,kumar2022estimation,mall2022broadband,wu2022accretion}, and the same way to study the second outburst in 2019 obtained a same range of $30^{\circ}-46^{\circ}$. The research of quasi-periodic oscillations (QPOs) suggested an inclination of $\sim 30^\circ$ \citep{liu2022transitions}. Due to the study of the inclination, the black hole mass, and the distance, the moderate spin of the BH ( $\sim 0.4$) is estimated using the continuum-fitting (CF) model \citep{wu2023moderate}.

Furthermore, the timing analysis of MAXI J1348-630 showed all of the type-A, type-B, and type-C QPOs appearing in the duration of the main outburst \citep{belloni2020time,alabarta2022variability,liu2020high,zhang2023type}. A type-C QPO with 0.9 Hz and a type-A QPO with 7 Hz were found in NICER and AstroSat observation \citep{jithesh2021broad}. Furthermore, the Comptonized region extends to 2300 km during the period when type-A QPO appeared \citep{zhang2023type}. For type-C QPO, the fractional rms amplitude increased in the energy band of 2-3 keV. The type-B QPO was discovered in $\sim 4.8 \rm Hz$ which may suggest the instability of the disk-jet structure \citep{liu2020high}, and about a decrease of half percent in fractional rms below 2 keV (at 1.5 keV) was first observed \citep{belloni2020time}. Besides, the radiative properties were explained by the two-component Comptonization model \citep{garcia2021two}. \cite{weng2021time} illustrated the time lag between the multi-temperature black body component produced by the disk and the power-law component made by the corona in the energy band between 0.5 to 80 keV. 

In this work, we provide a broad-band spectral view of MAXI J1348-630 during the hard and intermediate states observed by Insight-HXMT, concentrating on the non-thermal and reflection component evolution. In Section \ref{obdare}, the Insight-HXMT observations and data processes are introduced. In Section \ref{span}, we present the spectral analysis results and show the time lags between hard X-ray and radio emissions in Section \ref{tila}. In Section \ref{dis}, The physical implications of the spectral evolution and time lags of the multiwavelength bands are discussed. The brief summary and conclusion are given in Section \ref{con}.

\section{observations and data reduction}\label{obdare}
The first X-ray astronomy satellite of China, Insight Hard X-ray Modulation Telescope (Insight-HXMT, \citealt{zhang2020overview}), which launched on June 15 2017, observed the whole outburst of MAXI J1348-630 \citep{yatabe2019maxi} from 2019 January 27 to 2019 July 29 \citep{2019ATel12470....1C}. The entire energy range 1-250 keV of Insight-HXMT is based on three telescopes, namely, Low Energy X-ray telescope (LE) with 384 $\rm cm^{2}$ effective area from 1 keV to 15 keV \cite{chen2020low}, Medium Energy X-ray telescope (ME) with 952 $\rm cm^{2}$ from 5 keV to 30 keV \citep{cao2020medium}, and High Energy X-ray telescope (HE) with 5100 $\rm cm^{2}$ from 20 keV to 250 keV \citep{liu2020high}. The pile-up effects of Insight-HXMT are low enough to study the peak flux of this source ($< 1$ percent @ 18000 $\rm cts\, s^{-1}$, see \citealt{chen2020low}). There are no contaminating sources within the $4.3\times4.3\rm deg^{2}$ field of view (FOV) near MAXI J1348-630.

The Insight-HXMT Data was extracted to the spectra by using HXMTDAS (the Insight-HXMT Data Analysis software) v2.05 \footnote{http://hxmtweb.ihep.ac.cn/software.jhtml}. We jointly analyze three telescopes including 2 keV to 9 keV from LE telescope due to the low net count rate above 9 keV and the calibration uncertainties below 2 keV \citep{li2023orbit}, 11 keV to 21 keV and 24 keV to 30 keV from ME telescope due to the instrumental feature 21 keV to 24 keV and the low net count rate below 10 keV as well as above 30 keV \citep{cao2020medium}, the 30 keV to 100 keV from HE telescope because of the background noises at the same level with the net count rate below 30 keV and above 100 keV\citep{liu2020high}. To normalize these three telescopes, we insert three cross-normalization constants (the constant model in XSPEC), then, freeze the ME constant and let the LE and HE constants vary freely. We use the recommended criteria: the elevation angle $>$ 10 deg, the pointing offset angle $<$ 0.1 deg, the geomagnetic cutoff rigidity $>$ 8 GeV, and the good time intervals (GTIs) $>$ 300s away from the South Atlantic Anomaly (SAA), and generate the background files by LEBKGMAP for LE data, MEBKGMAP for ME data, HEBKGMAP for HE data to estimate the background count rates \citep{liao2020background}. The LELCGEN, MELCGEN, and HELCGEN tasks extract light curves from LE, ME, HE screened files. We used the ftgrouppha within FTOOLS to rebin the spectra data with a 40 minimum signal-to-noise ratio from 2.6 keV to 10 keV to promote precision at the iron emission line. A systematic error of 0.5\% was added to account for the calibration uncertainties. 

\section{Spectral analysis}\label{span}
The X-ray spectra were investigated by the XSPEC v12.13.0c\footnote{https://heasarc.gsfc.nasa.gov/xanadu/xspec/}. We fixed the galactic neutral hydrogen column density at $8.6\times 10^{21}\rm cm^{-2}$ in TBabs composition \citep{tominaga2020discovery,chakraborty2021nustar,saha2021multi,cangemi2022integral,zdziarski2022insight} of the all following model. The cross-normalization constants among LE, ME, and HE data were inserted in all models, freezing the ME constant as standard and letting the LE and HE constants vary freely. If the confidence level of parameters isn't mentioned in the following, it will be set to 90 percent to produce the error range and the statistic set to chi-squared statistic. 

\begin{figure}
    \centering
    \includegraphics[scale=0.8]{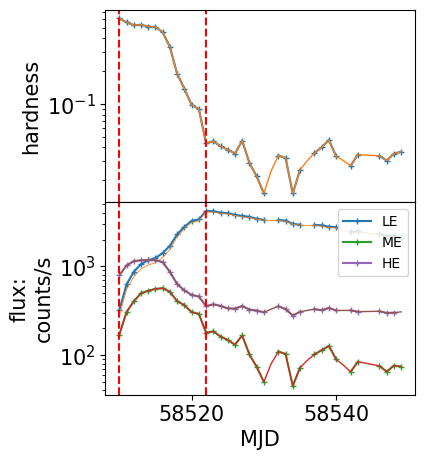}
    \caption{The hardness ratio and X-ray light curves of MAXI J1348-630 in units of count rates observed by Insight-HXMT. The top panel shows the hardness ratio between 6-10 keV and the 1-6 keV. The observational data that we investigate is between the two red lines (MJD 58510 to MJD 58519).}
    \label{LC}
\end{figure}

The main outburst of MAXI J1348-630 was observed by the Insight-HXMT from MJD 58510 to MJD 58693 and the HID of this source follows a 'q' pattern \citep{yatabe2019maxi,wu2023moderate}. In this research, we analyze the observational data in LHS and HIS illustrated in Fig. \ref{LC} between the two red lines. During LHS the hardness is near constant around 0.5-0.6, and the flux rises rapidly until MJD 58517, then the flux rises slowly, and the hardness drops from $\sim 0.4$ to 0.03 in HIS. In LHS and HIS, there are 28 times of observations from MJD 58510 to MJD 58522 provided by Insight-HXMT including all of the LE spectrum, ME spectrum, and HE spectrum, which are illustrated in Table \ref{ob} of the appendix, including observation ID, observation date, exposure time, and abbreviation of every observation. All the observational data we investigated have a hardness greater than 0.06, the HE data count rate greater than 400 counts/s and the ME data count rate greater than 200 counts/s (see Fig. \ref{LC}).

In the most spectra of LMBHXB, there exist three components: the multi-temperature black body from the disk, the cutoff power-law from the corona, the board iron line, and the hump over 20 keV from the reflection by disk \citep{remillard2006x}. At first, we use the phenomenological test models: power law and thermal Compton to fit the spectra. The first one is $$constant\ast Tbabs\ast (diskbb+cutoffpowerlaw),$$
which includes a cross-normalization constant among LE, ME, HE telescopes with the standard as the ME telescope, an interstellar absorption model Tbabs with \cite{wilms2000absorption} abundances and \cite{verner1996atomic} cross-section, a multicolor disk blackbody model diskbb \citep{mitsuda1984energy,makishima1986simultaneous} which is produced by the thin disk, and a power-law with high energy exponential cutoff component the component of which is produced by the corona.

\begin{figure}
    \centering
    \includegraphics[scale=0.6]{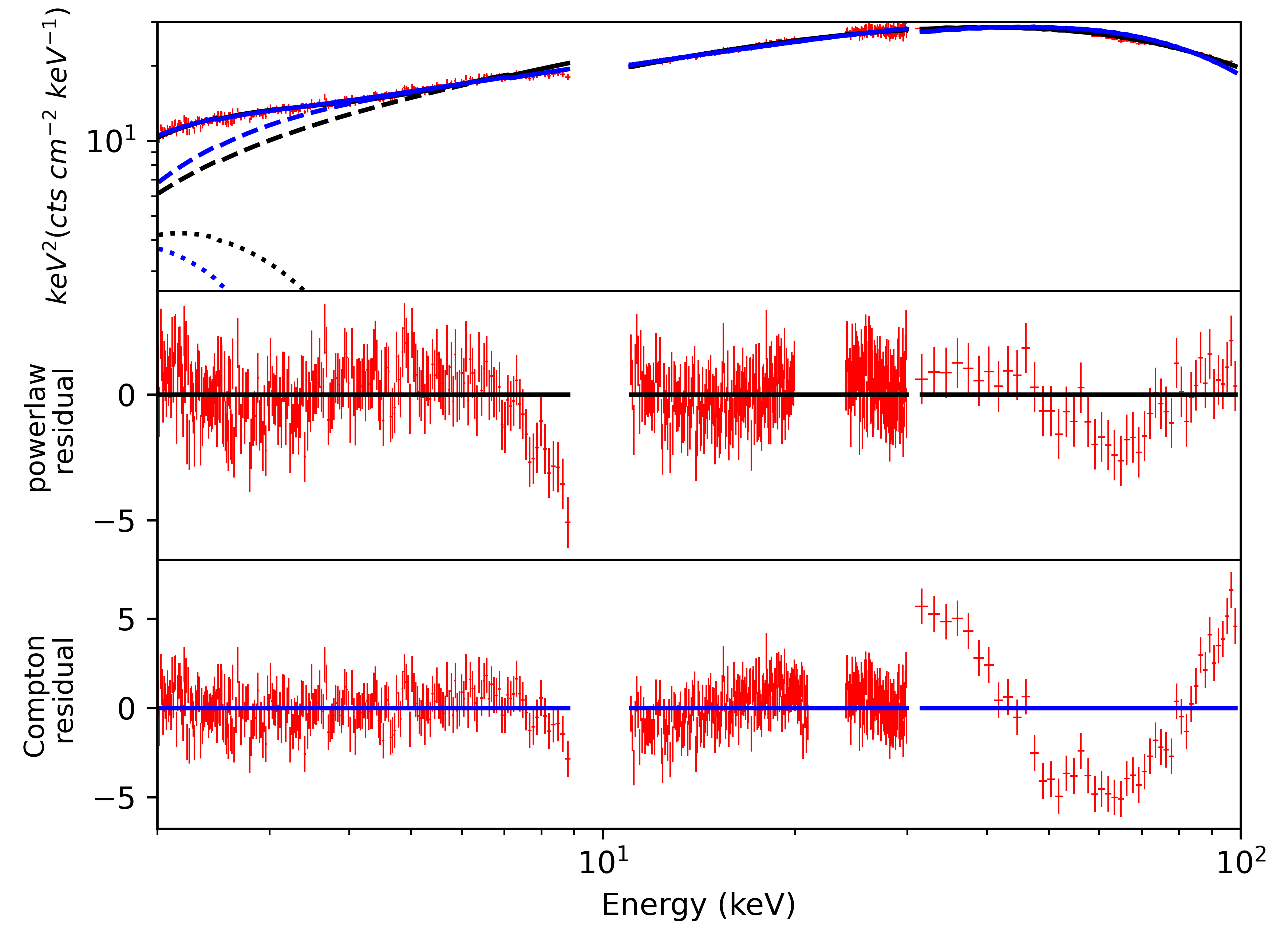}
    \caption{The example spectrum and residuals of observation 418 fitted with two test models. The blue line represents the results of the thermal Compton model (the dotted line for the additive component diskbb and the dashed line for the additive component nthcomp), the black line for the power law model (the dotted line for the component diskbb and the dashed line for the component cutoffpowerlaw), and the red point with the error bar for the data. }
    \label{cutoffpo}
\end{figure}

The thermal Compton model is:
$$constant \ast Tbabs \ast (diskbb+nthcomp),$$
where the nthcomp is the model describing the continuum shape component that is produced by the corona, this component is a thermal comptonization evolving a seed photon field which is part of disk flux scattering by corona \citep{zdziarski1996broad,zycki19991989}. In the entire studied period, the broad iron lines and the Compton hump could be noticed, which indicates the existence of a reflection component (see the example spectra and residuals in Fig. \ref{cutoffpo} and the dotted lines represent the disk component, the dashed lines for the corona component).

Therefore, in the following, we use the reflection model:
$$ constant \ast Tbabs \ast (diskbb+relxillCp) $$
to fit all the data for further studies. We use the relxillCp v2.0\footnote{https://www.sternwarte.uni-erlangen.de/~dauser/research/relxill/}. The relxillCp model could explain a broad iron line by assuming a primary thermal Compton spectrum \citep{dauser2014role,garcia2015estimating}. It should be pointed out that the corona source assumed by the lamppost model simplifies the corona as a point-like source staying beyond the black hole. Though this theoretical assumption is not realistic, the lamppost model still somehow approximates a realistic situation of the corona by some reasons, and can explain the spectral properties of BH X-ray binaries \citep{garcia2015estimating}. 

The part of flux from the corona could radiate on the accretion disk and be reflected by the accretion disk, then, it forms the reflected component that can be studied by the relxillCp model. The incident spectrum on the disk is assumed as a thermal Compton spectrum. In the fitting, we set the inclination at 30 degrees and the spin at 0.4 as a likely value, because the data can't estimate these two parameters with other parameters at the same time and these two parameters of this source have been studied well in other research \citep{wu2023moderate,chakraborty2021nustar,carotenuto2021black}. \cite{chakraborty2021nustar} pointed out that the disk density of this source exceeds $10^{20}\rm cm^{-3}$. Due to the limitation of the maximum value $10^{20}\rm cm^{-3}$, we set the density of the disk as $10^{20}\rm cm^{-3}$. The abundance of the iron $A_{fe}$ can't be constrained around 1, and the $relxillCp$ generally gives a systematically large iron abundance due to the limitation of the model \citep{chakraborty2021nustar}, so we fix $A_{fe}=3.5$ in units of the solar abundance. The outer edge of the accretion disk and the emissivity are fixed at a maximum value of 400 $r_{g}$ and 3 respectively. 

The fitted parameters are shown together in Table \ref{Rre}. The relxillCp parameters include $r_{in}$ for the inner edge of the accretion disk in the unity of the ISCO, $\Gamma_{rel}$ for the photon index, $\xi$ for the ionization parameter in logarithmic units of $\rm erg\ cm\ s^{-1}$, $kT_{e}$ for the temperature of the high energy electron (corona) in the unity of keV, $n_{disk}$ for the density of the accretion disk in logarithmic units of $cm^{-3}$, $F_{fr}$ for the reflection friction (ratio of intensity emitted to the disk to escaping to infinity), $N_{re}$ for an incident flux to the disk \citep{dauser2016normalizing}. We set the upper limit of $r_{in}$ as 20 ISCO.

\begin{table*}

		\centering
		\caption{Spectral fitting results for the reflection model : constant $\ast$ Tbabs $\ast$ (diskbb+relxillCp)}
		\label{Rre}
		\begin{tabular}{lccccccccr} 
			\hline
			 & diskbb  & &relxillCp &&&& &  \\
			Obs.  &  $T_{in}$ & $N_{diskbb}$ & $r_{in}$ & $\Gamma_{rel}$ & $log\xi$ & $kT_{e}$ & $F_{fr}$& $N_{re}$\\ 
   &  (keV) &  & (ISCO) &  & $(\rm erg\ cm\ s^{-1})$ & (keV)  & & \\ \hline

    101 &     $0.65_{-0.10}^{+0.02}$ & $1000 _{-1000 }^{+1000 }$ & $25^{+1}_{-18}$ & $1.46_{-0.01}^{+0.05}$  & $4.25_{-0.16}^{+0.05}$  & $75_{-24}^{+1}$ & $0.54_{-0.25}^{+0.03}$  & $0.05_{-0.01}^{+0.01}$   \\
    201 &     $0.56_{-0.04}^{+0.04}$ & $3000 _{-1000 }^{+1000 }$ & $15^{+9}_{-10}$ & $1.51_{-0.02}^{+0.01}$  & $4.14_{-0.13}^{+0.17}$  & $45_{-6}^{+4}$ & $0.56_{-0.13}^{+0.11}$  & $0.08_{-0.01}^{+0.01}$   \\
    301 &     $0.50_{-0.04}^{+0.02}$ & $8000 _{-1000 }^{+3000 }$ & $18^{+6}_{-11}$ & $1.54_{-0.01}^{+0.01}$  & $4.11_{-0.12}^{+0.09}$  & $44_{-3}^{+4}$ & $0.67_{-0.10}^{+0.14}$  & $0.10_{-0.01}^{+0.01}$   \\
    401 &     $0.54_{-0.03}^{+0.03}$ & $7000 _{-1000 }^{+2000 }$ & $21^{+3}_{-14}$ & $1.56_{-0.01}^{+0.01}$  & $4.14_{-0.10}^{+0.12}$  & $44_{-5}^{+3}$ & $0.83_{-0.18}^{+0.10}$  & $0.10_{-0.01}^{+0.01}$   \\
    403 &     $0.57_{-0.03}^{+0.02}$ & $7000 _{-1000 }^{+2000 }$ & $22^{+3}_{-14}$ & $1.56_{-0.01}^{+0.02}$  & $4.33_{-0.11}^{+0.05}$  & $39_{-4}^{+6}$ & $0.67_{-0.16}^{+0.22}$  & $0.11_{-0.01}^{+0.01}$   \\
    404 &     $0.50_{-0.04}^{+0.04}$ & $12000_{-4000 }^{+6000 }$ & $21^{+4}_{-13}$ & $1.59_{-0.02}^{+0.01}$  & $4.12_{-0.13}^{+0.13}$  & $48_{-5}^{+15}$ & $0.83_{-0.13}^{+0.28}$  & $0.10_{-0.01}^{+0.01}$   \\
    405 &     $0.54_{-0.04}^{+0.04}$ & $9000 _{-3000 }^{+4000 }$ & $15^{+9}_{-10}$ & $1.56_{-0.02}^{+0.01}$  & $4.32_{-0.10}^{+0.08}$  & $40_{-3}^{+5}$ & $0.73_{-0.11}^{+0.23}$  & $0.10_{-0.01}^{+0.01}$   \\
    407 &     $0.53_{-0.05}^{+0.05}$ & $9000 _{-3000 }^{+6000 }$ & $6^{+18}_{-1}$ & $1.56_{-0.01}^{+0.02}$  & $4.11_{-0.10}^{+0.31}$  & $45_{-13}^{+3}$ & $1.04_{-0.54}^{+0.06}$  & $0.09_{-0.01}^{+0.03}$   \\
    408 &     $0.58_{-0.04}^{+0.01}$ & $7000 _{-1000 }^{+2000 }$ & $5^{+19}_{-1}$ & $1.55_{-0.01}^{+0.02}$  & $4.36_{-0.22}^{+0.01}$  & $38_{-2}^{+6}$ & $0.75_{-0.08}^{+0.21}$  & $0.10_{-0.01}^{+0.01}$   \\
    410 &     $0.53_{-0.04}^{+0.03}$ & $11000_{-2000 }^{+4000 }$ & $15^{+9}_{-12}$ & $1.57_{-0.01}^{+0.02}$  & $4.30_{-0.17}^{+0.09}$  & $39_{-3}^{+7}$ & $0.74_{-0.13}^{+0.23}$  & $0.11_{-0.01}^{+0.01}$   \\
    411 &     $0.53_{-0.03}^{+0.03}$ & $9000 _{-2000 }^{+3000 }$ & $22^{+2}_{-15}$ & $1.58_{-0.01}^{+0.01}$  & $4.14_{-0.10}^{+0.12}$  & $43_{-5}^{+4}$ & $0.85_{-0.17}^{+0.13}$  & $0.10_{-0.01}^{+0.01}$   \\
    412 &     $0.54_{-0.03}^{+0.03}$ & $10000_{-2000 }^{+3000 }$ & $22^{+2}_{-17}$ & $1.58_{-0.01}^{+0.01}$  & $4.19_{-0.12}^{+0.12}$  & $43_{-6}^{+4}$ & $0.89_{-0.20}^{+0.15}$  & $0.10_{-0.01}^{+0.01}$   \\
    415 &     $0.48_{-0.06}^{+0.03}$ & $18000_{-6000 }^{+19000}$ & $3^{+20}_{-1}$ & $1.59_{-0.01}^{+0.03}$  & $4.22_{-0.14}^{+0.04}$  & $45_{-8}^{+3}$ & $0.93_{-0.33}^{+0.09}$  & $0.10_{-0.01}^{+0.02}$   \\
    416 &     $0.45_{-0.04}^{+0.02}$ & $20000_{-4000 }^{+13000}$ & $8^{+16}_{-3}$ & $1.59_{-0.01}^{+0.02}$  & $4.04_{-0.12}^{+0.08}$  & $45_{-5}^{+9}$ & $0.98_{-0.23}^{+0.19}$  & $0.10_{-0.01}^{+0.01}$   \\
    417 &     $0.51_{-0.08}^{+0.01}$ & $12000_{-2000 }^{+11000}$ & $3^{+20}_{-1}$ & $1.58_{-0.01}^{+0.03}$  & $4.19_{-0.28}^{+0.12}$  & $39_{-3}^{+8}$ & $0.83_{-0.16}^{+0.26}$  & $0.11_{-0.01}^{+0.01}$   \\
    418 &     $0.41_{-0.07}^{+0.03}$ & $38000_{-14000}^{+75000}$ & $14^{+10}_{-11}$ & $1.61_{-0.01}^{+0.02}$  & $4.02_{-0.11}^{+0.08}$  & $46_{-5}^{+10}$ & $0.95_{-0.20}^{+0.23}$  & $0.10_{-0.01}^{+0.01}$   \\
    419 &     $0.40_{-0.04}^{+0.05}$ & $35000_{-16000}^{+27000}$ & $12^{+11}_{-9}$ & $1.62_{-0.01}^{+0.03}$  & $3.92_{-0.09}^{+0.19}$  & $48_{-12}^{+1}$ & $0.99_{-0.45}^{+0.02}$  & $0.10_{-0.01}^{+0.03}$   \\
    420 &     $0.46_{-0.10}^{+0.05}$ & $20000_{-7000 }^{+63000}$ & $20^{+4}_{-16}$ & $1.62_{-0.02}^{+0.03}$  & $4.11_{-0.16}^{+0.17}$  & $40_{-4}^{+7}$ & $0.73_{-0.18}^{+0.21}$  & $0.12_{-0.01}^{+0.02}$   \\
    501 &     $0.42_{-0.09}^{+0.05}$ & $39000_{-19000}^{+17000}$ & $5^{+18}_{-3}$ & $1.63_{-0.01}^{+0.03}$  & $4.06_{-0.17}^{+0.08}$  & $47_{-4}^{+14}$ & $1.05_{-0.16}^{+0.28}$  & $0.10_{-0.01}^{+0.01}$   \\
    601 &     $0.39_{-0.05}^{+0.03}$ & $75000_{-23000}^{+95000}$ & $7^{+5}_{-3}$ & $1.73_{-0.02}^{+0.03}$  & $3.94_{-0.08}^{+0.08}$  & $94_{-16}^{+26}$ & $2.10_{-0.30}^{+0.48}$  & $0.06_{-0.01}^{+0.01}$   \\
    602 &     $0.41_{-0.03}^{+0.01}$ & $68000_{-10000}^{+41000}$ & $6^{+17}_{-1}$ & $1.76_{-0.01}^{+0.02}$  & $3.86_{-0.07}^{+0.03}$  & $85_{-5}^{+48}$ & $1.91_{-0.06}^{+0.58}$  & $0.07_{-0.01}^{+0.01}$   \\
    603 &     $0.46_{-0.06}^{+0.01}$ & $51000_{-4000 }^{+44000}$ & $1.38 ^{+10}_{-0.21 }$ & $1.71_{-0.01}^{+0.05}$  & $4.14_{-0.28}^{+0.04}$  & $91_{-1}^{+26}$ & $2.33_{-0.16}^{+0.42}$  & $0.05_{-0.01}^{+0.01}$   \\
    701 &     $0.55_{-0.01}^{+0.03}$ & $31000_{-8000 }^{+5000 }$ & $1.01 ^{+0.25 }_{-0.01 }$ & $1.97_{-0.02}^{+0.05}$  & $3.47_{-0.24}^{+0.15}$  & $124_{-52}^{+6}$ & $1.00_{-0.36}^{+0.01}$  & $0.10_{-0.01}^{+0.03}$   \\
    702 &     $0.61_{-0.01}^{+0.01}$ & $23000_{-2000 }^{+3000 }$ & $1.00 ^{+0.12 }_{-0.01 }$ & $2.03_{-0.04}^{+0.02}$  & $3.30_{-0.11}^{+0.18}$  & $123_{-33}^{+6}$ & $1.03_{-0.22}^{+0.14}$  & $0.10_{-0.02}^{+0.02}$   \\
    804 &     $0.66_{-0.01}^{+0.01}$ & $22000_{-1000 }^{+1000 }$ & $1.00 ^{+0.05 }_{-0.01 }$ & $2.07_{-0.01}^{+0.02}$  & $3.30_{-0.08}^{+0.02}$  & $124_{-14}^{+8}$ & $1.07_{-0.21}^{+0.05}$  & $0.10_{-0.01}^{+0.01}$   \\
    904 &     $0.70_{-0.01}^{+0.01}$ & $20000_{-1000 }^{+2000 }$ & $1.00 ^{+0.15 }_{-0.01 }$ & $2.16_{-0.02}^{+0.02}$  & $3.01_{-0.08}^{+0.13}$  & $124_{-14}^{+7}$ & $0.53_{-0.13}^{+0.13}$  & $0.13_{-0.02}^{+0.02}$   \\
    1101 &    $0.76_{-0.01}^{+0.01}$ & $24000_{-1000 }^{+1000 }$ & $1.00 ^{+0.05 }_{-0.01 }$ & $2.21_{-0.01}^{+0.01}$  & $3.00_{-0.03}^{+0.03}$  & $124_{-10}^{+8}$ & $0.90_{-0.14}^{+0.14}$  & $0.08_{-0.01}^{+0.01}$   \\
                \hline
			 &constant & & & &&constant&&&\\
			Obs.  & $C_{LE}$ & $C_{HE}$ & $\nu/\chi^{2}$& $\chi^{2}_{\nu}$  & Obs.  & $C_{LE}$ & $C_{HE}$ & $\nu/\chi^{2}$& $\chi^{2}_{\nu}$  \\
			\hline
    101 &  $0.95_{-0.04}^{+0.04}$  & $1.01_{-0.01}^{+0.02}$  & 437.23/480 &0.91   &    417 &   $0.99_{-0.04}^{+0.02}$  & $0.99_{-0.01}^{+0.01}$  & 481.51/529 &0.91     \\ 
    201 &  $1.00_{-0.04}^{+0.03}$  & $0.99_{-0.01}^{+0.01}$  & 540.95/562 &0.96   &    418 &   $0.96_{-0.04}^{+0.02}$  & $1.00_{-0.01}^{+0.01}$  & 429.57/532 &0.81     \\ 
    301 &  $0.95_{-0.02}^{+0.03}$  & $0.98_{-0.01}^{+0.01}$  & 593.54/666 &0.89   &    419 &   $0.97_{-0.04}^{+0.02}$  & $1.01_{-0.01}^{+0.02}$  & 542.33/562 &0.97     \\ 
    401 &  $0.97_{-0.02}^{+0.02}$  & $0.98_{-0.01}^{+0.01}$  & 769.44/763 &1.01   &    420 &   $0.97_{-0.06}^{+0.03}$  & $1.01_{-0.01}^{+0.01}$  & 352.56/411 &0.86     \\ 
    403 &  $0.97_{-0.02}^{+0.03}$  & $0.99_{-0.01}^{+0.01}$  & 604.68/601 &1.01   &    501 &   $1.00_{-0.07}^{+0.01}$  & $1.00_{-0.01}^{+0.01}$  & 427.20/414 &1.03     \\ 
    404 &  $0.93_{-0.02}^{+0.03}$  & $0.99_{-0.01}^{+0.01}$  & 548.55/555 &0.99   &    601 &   $0.95_{-0.05}^{+0.03}$  & $1.02_{-0.02}^{+0.02}$  & 403.43/467 &0.86     \\ 
    405 &  $0.96_{-0.02}^{+0.05}$  & $0.98_{-0.01}^{+0.01}$  & 457.02/463 &0.99   &    602 &   $0.91_{-0.03}^{+0.02}$  & $0.98_{-0.01}^{+0.01}$  & 552.27/572 &0.97     \\ 
    407 &  $1.01_{-0.04}^{+0.05}$  & $1.01_{-0.01}^{+0.01}$  & 460.92/463 &1.00   &    603 &   $0.95_{-0.06}^{+0.02}$  & $0.97_{-0.01}^{+0.02}$  & 462.68/491 &0.94     \\ 
    408 &  $0.98_{-0.03}^{+0.02}$  & $0.97_{-0.01}^{+0.01}$  & 586.71/616 &0.95   &    701 &   $1.03_{-0.04}^{+0.03}$  & $0.98_{-0.02}^{+0.03}$  & 562.59/572 &0.98     \\ 
    410 &  $1.01_{-0.04}^{+0.03}$  & $0.99_{-0.01}^{+0.01}$  & 515.98/501 &1.03   &    702 &   $0.95_{-0.01}^{+0.03}$  & $0.98_{-0.01}^{+0.01}$  & 625.64/612 &1.02     \\ 
    411 &  $0.98_{-0.02}^{+0.03}$  & $0.99_{-0.01}^{+0.01}$  & 690.55/694 &1.00   &    804 &   $0.96_{-0.02}^{+0.03}$  & $1.00_{-0.02}^{+0.01}$  & 723.51/663 &1.09    \\ 
    412 &  $0.98_{-0.03}^{+0.03}$  & $0.98_{-0.01}^{+0.01}$  & 538.60/579 &0.93   &    904 &   $1.09_{-0.04}^{+0.04}$  & $0.97_{-0.02}^{+0.02}$  & 577.30/509 &1.13    \\ 
    415 &  $0.94_{-0.04}^{+0.03}$  & $0.99_{-0.01}^{+0.01}$  & 432.65/471 &0.92   &    1101 &  $1.05_{-0.03}^{+0.03}$  & $1.04_{-0.01}^{+0.03}$  & 866.20/688 &1.26     \\ 
    416 &  $0.95_{-0.03}^{+0.02}$  & $0.97_{-0.01}^{+0.01}$  & 605.82/642 &0.94        \\ \hline

\end{tabular}
\end{table*}

\begin{figure}
    \centering
    \includegraphics[scale=0.6]{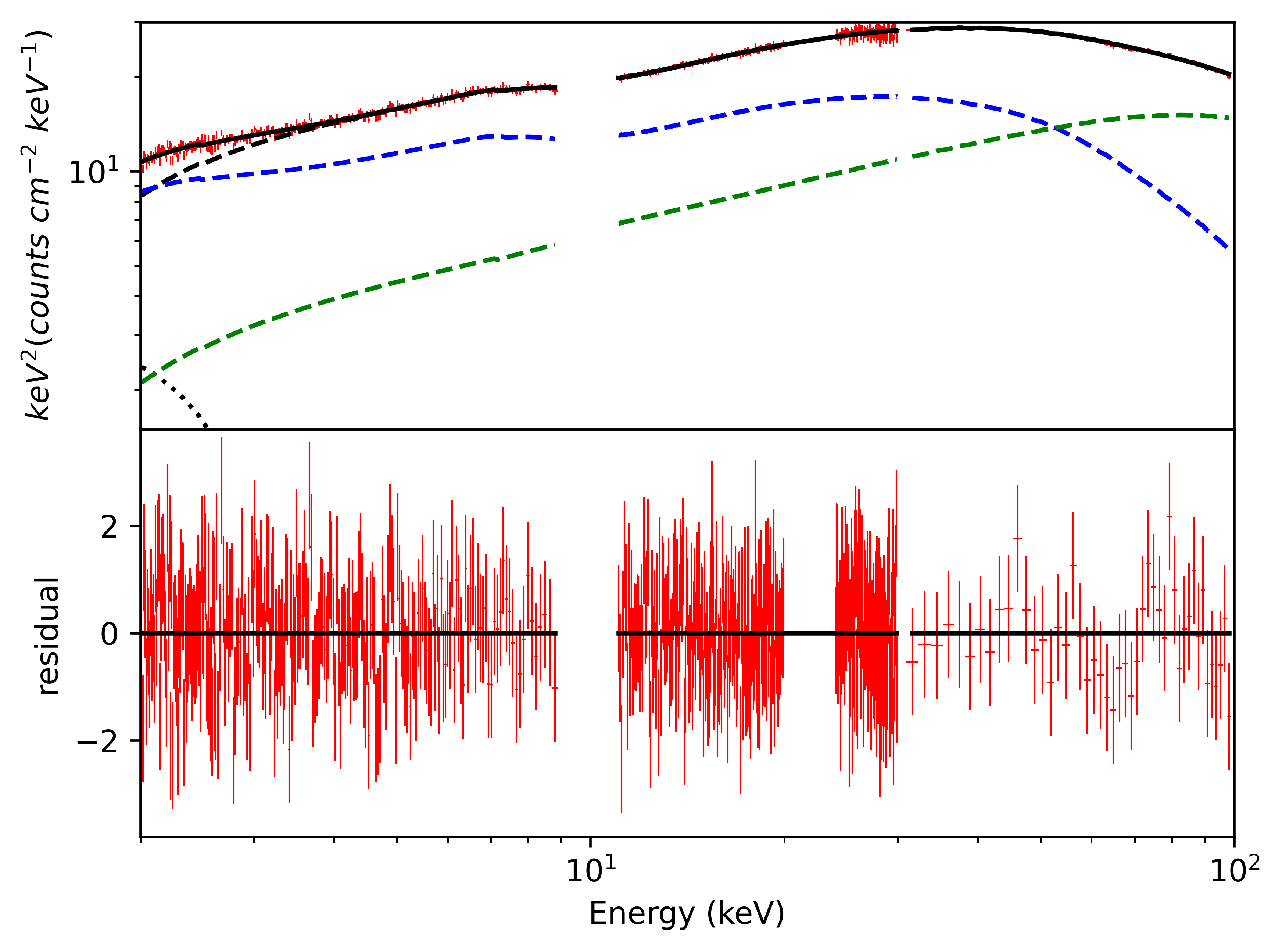}
    \caption{The example spectrum and residual of observation 418 with the reflection model. The solid line represents the model, the dotted line for the additive component diskbb and the dashed line for the component relxillCp, the blue line for the component which is reflected by the disk and the green line for the component reaching the observer directly and the red points with error bars for the data.}
    \label{ref}
\end{figure}

\begin{figure*}
    \centering
    \includegraphics[scale=0.275]{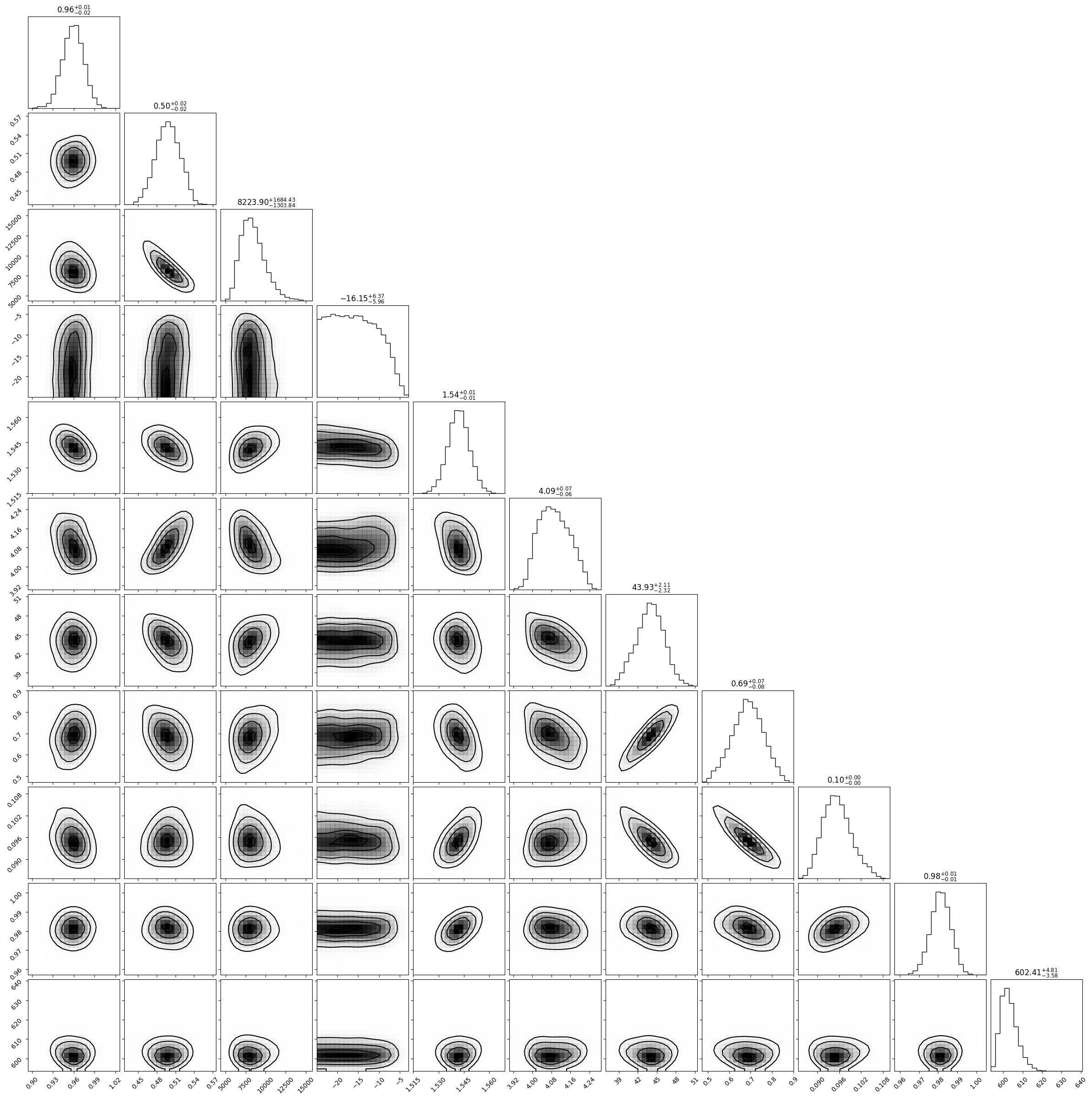}
    \caption{This figure shows the contour diagram for observation 404. There is no apparent parameter degeneration between these parameters (the contour levels plot as 0.5, 1, 1.5, 2 $\sigma$).}
    \label{recont5}
\end{figure*}

The reflection model can well fit the spectra with reduced $\chi^{2}$ around 1 and there is no structure on the residuals of all observations (the example spectrum and residual of observation 418 are shown in Fig. \ref{ref}), besides, the cross normalization of three telescopes (LE, ME, HE) is at 0.95 to 1.05. The MCMC method with the Metropolis-Hastings algorithm is used to calculate the results and uncertainties by applying $7\times 10^{5}$ burning steps and $1.5\times 10^{6}$ steps length with 10 walkers (the example contour diagrams are illustrated in Fig. \ref{recont5}). 

\begin{figure}
    \centering
    \includegraphics[scale=0.5]{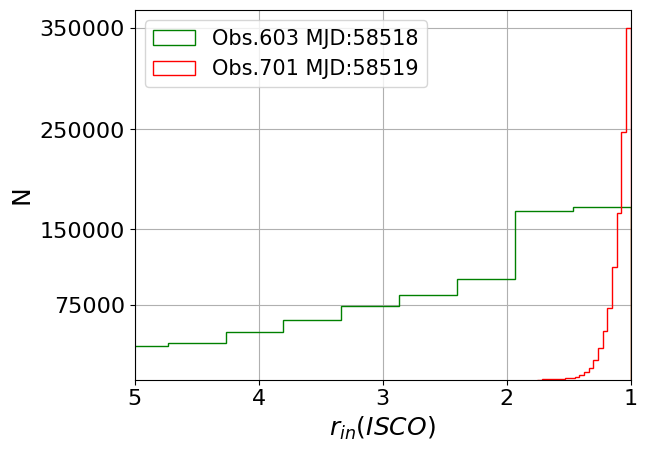}
    \caption{This figure shows the MCMC results of the accretion disk inner edge parameter $r_{in}$ in Obs. 603 at MJD 58518 for green histogram and Obs. 701 at MJD 58519 for red histogram.}
    \label{rin}
\end{figure}

\begin{figure}
    \centering
    \includegraphics[scale=0.5]{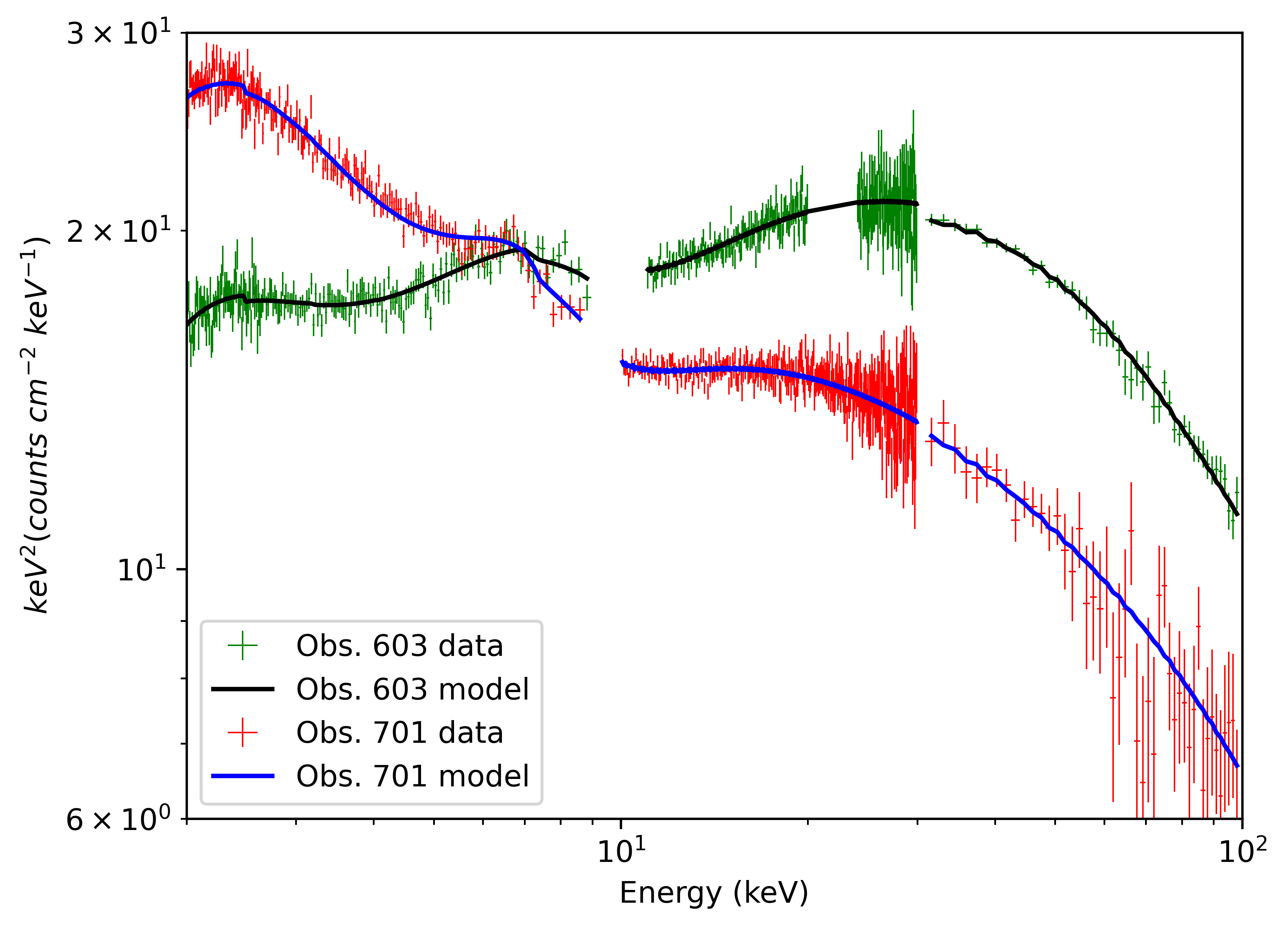}
    \caption{This figure shows the spectral data for Obs. 603 at MJD 58518  (green points) and Obs. 701 at MJD 58519 (red points).}
    \label{compare}
\end{figure}

The results of the reflection model indicate that after observation 603 (at the observation 603, $r_{in}$ is very close to the ISCO) at MJD 58518, the inner edge of the accretion disk reaches the ISCO, and before that date $r_{in}$ has the low limit around 5 ISCO as illustrated in the $r_{in}$ row of Table \ref{Rre}. Fig. \ref{rin} shows the MCMC results after marginalizing the other parameters, and illustrates that in Obs. 603 (MJD 58518), $r_{in}$ presses within 2 ISCO and in Obs. 701 (MJD 58519), $r_{in}$ reaches ISCO. The spectra of these two observations shown in Fig. \ref{compare} also implied that when the accretion disk reaches the ISCO, the relatively soft flux (below 7 keV) becomes the dominant compared to the flux beyond 7 keV, and the flux at the energy band around iron emission line (near 7 keV) is invariable.

\begin{figure}
    \centering
    \includegraphics[scale=0.6]{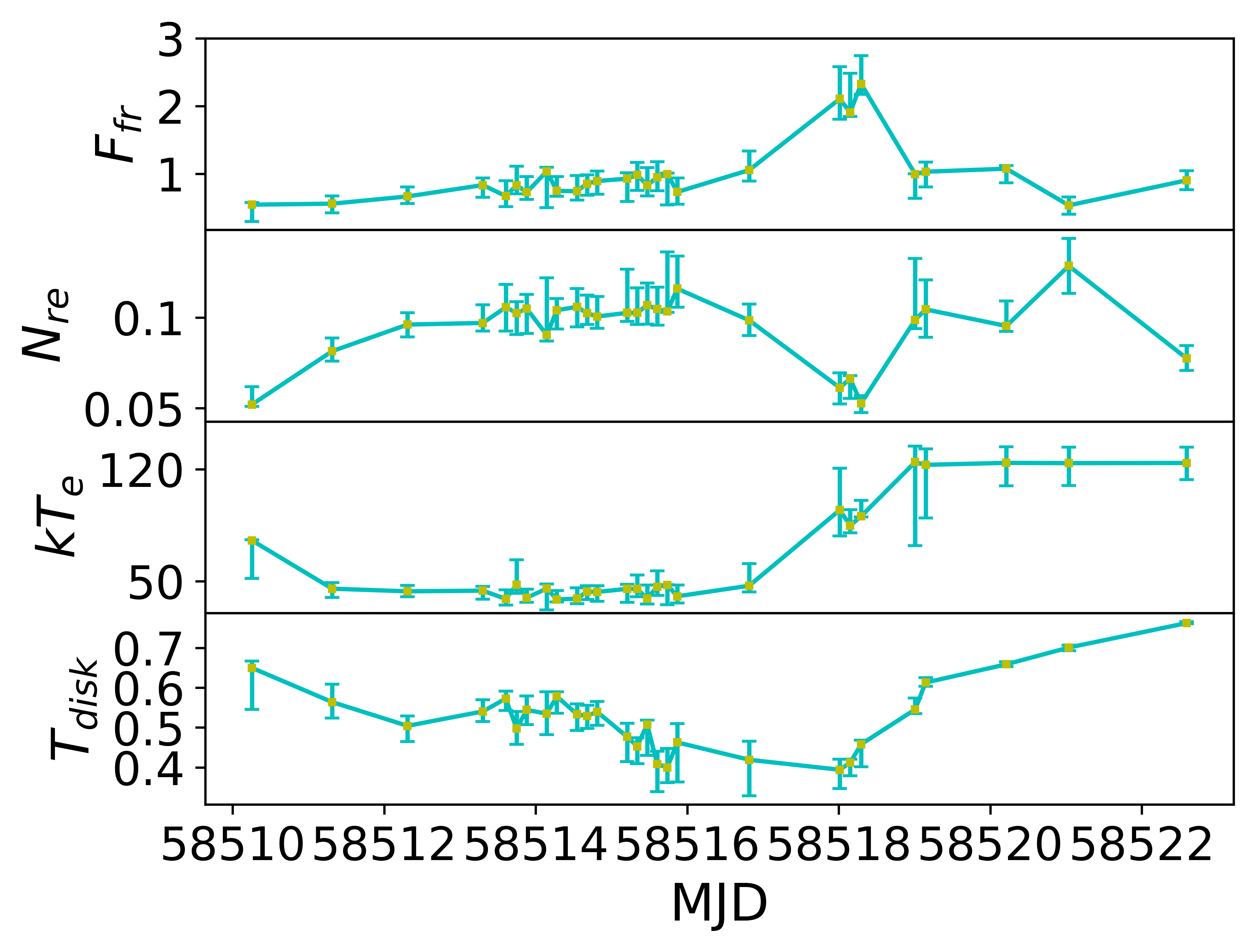}
    \caption{The evolution of the spectral parameters: the reflection fraction $F_{fr}$, the reflection normalization $N_{re}$, the temperature of the high energy electron (corona) $kT_{e}$ in unity of keV, and the temperature of the accretion disk $T_{in}$ in unity of keV. }
    \label{refpa}
\end{figure}

\begin{figure}
    \centering
    \includegraphics[scale=0.7]{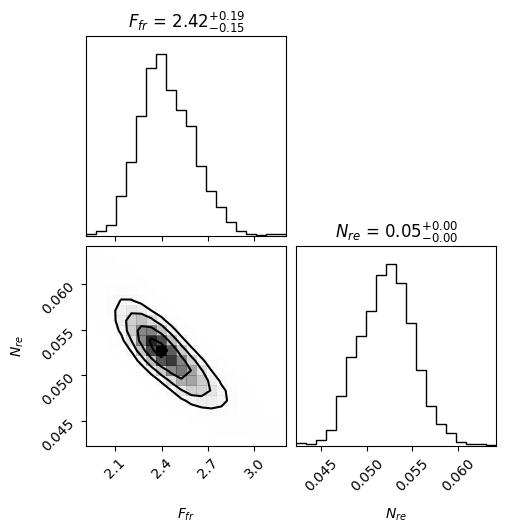}
    \caption{The contour diagram of the reflection fraction $F_{re}$ and the reflection normalization $N_{re}$ for observation 603. There is no apparent parameter degeneration between these parameters (the contour levels plot as 0.5, 1, 1.5, 2 $\sigma$).}
    \label{Frecont}
\end{figure}

The top 2 panels of Fig. \ref{refpa} illustrate the reflection fraction $F_{fr}$ (ratio of intensity emitted to the disk and to escaping to infinity) and the normalization of reflection $N_{re}$ for the incident spectrum flux ($\int^{1MeV}_{0.1keV} F_{E}dE=10^{20}\frac{n\xi}{4\pi}$, where $F_{E}(E)$ is the flux and $n$ fix at $10^{15}\rm cm^{-3}$, $\xi=1\rm erg\ cm\ s^{-1}$, see \citealt{dauser2016normalizing}). The $F_{fr}$ is generally below 1, except MJD 58517, 58518 when $F_{fr}$ rises in MJD 58517 and reaches over 2, and then decays to less than 1 during HIS. For the reflection normalization $N_{re}$, at the beginning of the outburst, the reflection normalization increases from 0.05 to 0.1 from MJD 58510 to 58512 and drops down to 0.05 around MJD 58518, rises above 0.1 in HIS. The MCMC results show that there is no apparent parameter degeneration between $F_{re}$ and $N_{re}$ (Fig. \ref{Frecont}). The change in $F_{fr}$ and $N_{re}$ implies the corona evolution. The temperature of corona $kT_{e}$ in the range of $\sim 50 - 100$ keV, and in LHS the corona cools down to 50 keV, when the state to IS, the corona is heated back to 100 keV.

\begin{figure}
    \centering
    \includegraphics[scale=0.5]{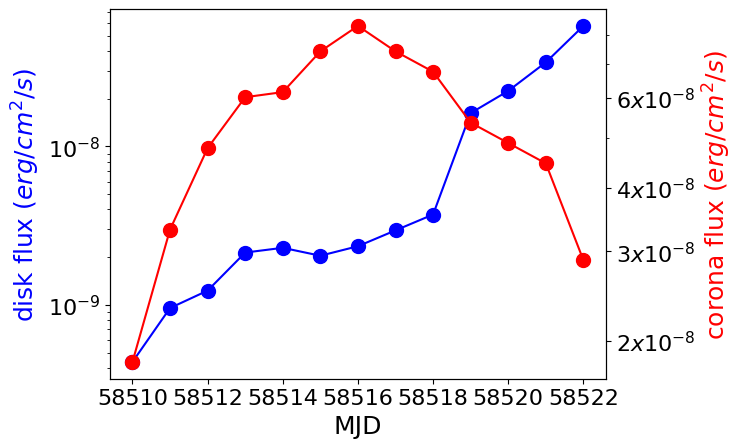}
    \caption{The unabsorbed flux evolution of the disk and corona components from the beginning of the outburst to the end of HIS. The unabsorbed disk flux is obtained from the integral of diskbb in 2 keV to 100 keV corresponding to the blue points using the left Y-axis. The red points correspond to the unabsorbed corona flux, which is calculated by relxillCp in the reflection model in MJD order using the right Y-axis. }
    \label{relflux}
\end{figure}

\begin{figure}
    \centering
    \includegraphics[scale=0.5]{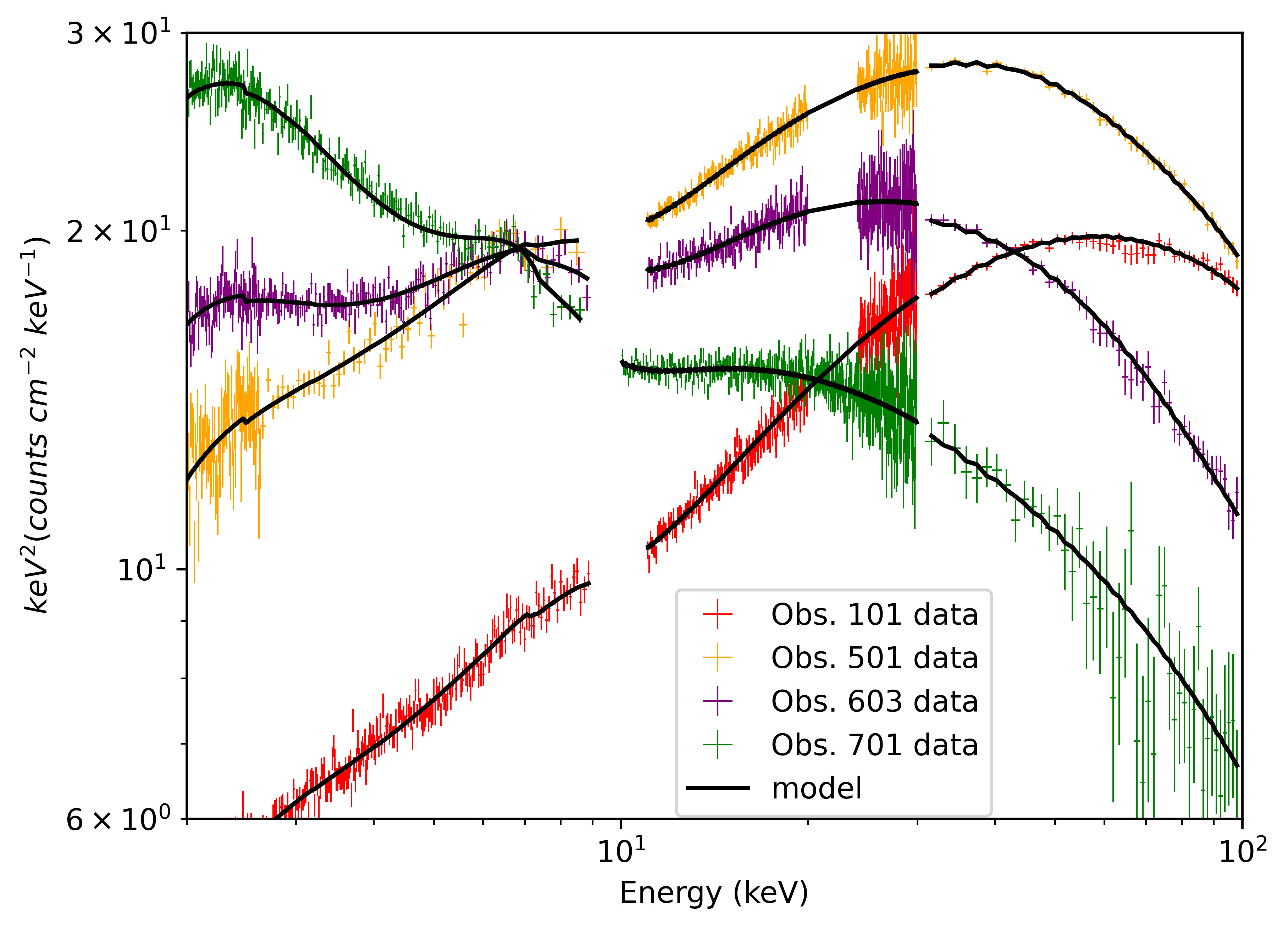}
    \caption{This figure shows the spectral examples and variations in different states. The red points represent the initial HS (Obs. 101), the orange points for the peak in the HS (Obs. 501), the purple points for the HIS (Obs. 603), the green points for the SIS (Obs. 701).}
    \label{state}
\end{figure}

The unabsorbed fluxes in both disk and corona components are calculated which are obtained from the integral of diskbb and relxillCp in 2 keV to 100 keV as shown in Fig. \ref{relflux}. The disk flux continued to increase during the observations, from $2\times10^{-9}\rm erg/cm^{2}/s$ to $4\times10^{-9}\rm erg/cm^{2}/s$ during LHS, and rise fast during HIS from $\sim 10^{-8}\rm erg/cm^{2}/s$ to $6\times10^{-8}\rm erg/cm^{2}/s$ within 3 days. At the beginning of the outburst, the corona flux around $2\times 0^{-8}\rm erg/cm^{2}/s$ is one magnitude higher than the disk flux, and increases faster than that of the disk flux during the HIS. The corona component shows the peak flux around $8\times10^{-8}\rm erg/cm^{2}/s$, and drops with the state evolving from LHS to HIS. At the end of HIS, the corona flux is about half of the disk flux. We also present the spectral changes in the state transitions in Fig. \ref{state}. With the state transitions, the flux below the 7 keV continued to increase, but the flux above 7 keV increased to a peak, then returned to drop, and the hump around 30 - 50 keV disappeared.


\section{time lag of the radio peak}\label{tila}

\begin{figure}
    \centering
    \includegraphics[scale=0.5]{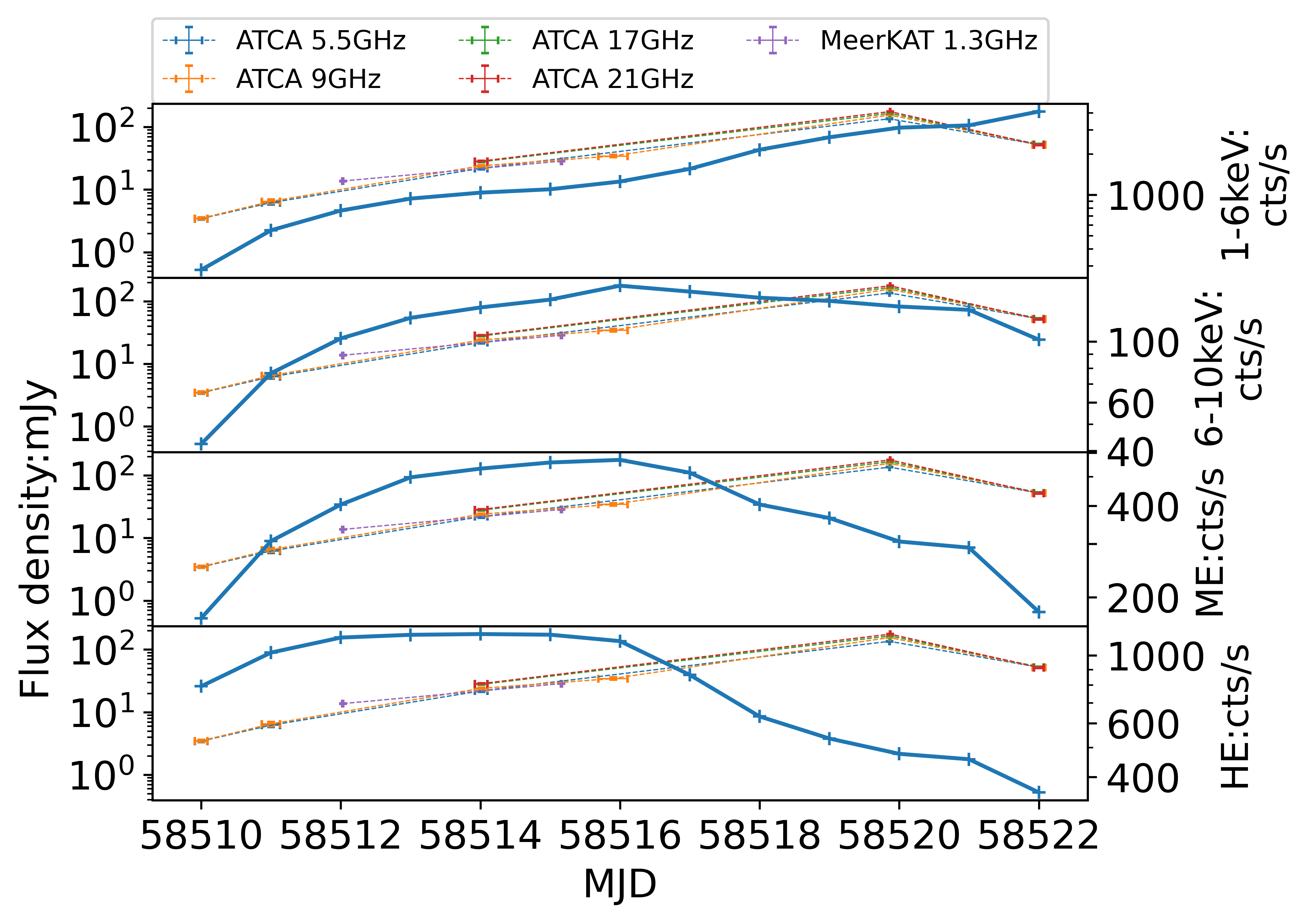}
    \caption{The light curves of the radio flux density compared to X-ray ones in different energy bands: the top panel for the 1-6 keV light curve, the second panel for 6-10 keV, the third panel for ME telescope (11 keV - 30 keV), the bottom panel for HE telescope (30 keV - 100 keV). The ratio flux density is observed by ATCA and MeerKAT in different frequencies (ATCA 5.5GHz for green dotted line; ATCA 9GHz for orange dotted line; ATCA 17GHz for light green dotted line; ATCA 21GHz for red dotted line; MeerKAT 1.3GHz for purple dotted line). }
    \label{radio}
\end{figure}

The MeerKAT and ATCA \citep{jonas2016meerkat,carotenuto2020meerkat} observed the source during the entire period we studied \citep{carotenuto2021black}, which illustrated a radio peak at MJD 58520. In Fig. \ref{radio}, we compare the light curves of the radio flux density and X-rays in different energy bands. The light curve of the soft X-rays from 1 - 6 keV didn't reach the peak flux until MJD 58522, then the disk emission continued to increase during the observations (also see Fig. \ref{relflux}). The harder X-ray photon emission above 6 keV reached the peak in the order from high energy band to low energy band. The 6-10 keV photon flux reaches a peak at around MJD 58516, and decreases slowly. The photon flux 11-30 keV (collected by ME telescopes) rises rapidly from the beginning of the outburst and reaches the peak at around MJD 58515-58516, and then drops to one-third within a week. For the 30-100 keV photons, the flux rises smoothly, reaches a peak phase around MJD 58512-58514, then decreases to one-third within six days. The radio flux density evolution presents a similar pattern in the radio bands from 1.3 GHz to 21 GHz observed by MeerKAT and ATCA, and the flux density increases smoothly from 3 mJy to the peak of over 100 mJy at MJD 58520, and then drops. Thus the radio emission shows a time lag compared with hard X-ray emissions. 

\begin{figure}
    \centering
    \includegraphics[scale=0.5]{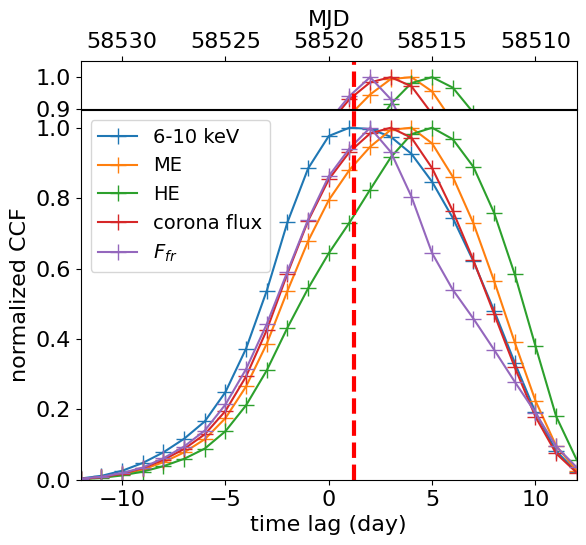}
    \caption{The cross-correlation function (CCF normalized with the max bin) between the radio flux density and X-ray light curves (6 keV - 10 keV for blue line; ME telescope for orange line; HE telescope for green line; corona flux for red solid line; the $F_{fr}$ curve for purple line). The dashed red line indicates the date when $r_{in}$ reaches ISCO.}
    \label{ccf}
\end{figure}

We used the cross-correlation function (CCF) to investigate the time lags of radio emission (shown in Fig. \ref{ccf}). Due to the radio flux density peak after all other curves, we set the other curves as zero point to study the radio time lag, thus, the bottom X-axis of Fig. \ref{ccf} set as the radio peak time lag compared to other band curves in unity of day and the top X-axis of Fig. \ref{ccf} represents MJD (the radio peak at MJD 58520 is aligned with the zero-day time lag). We normalized the CCF with the max bin of CCF results. The results of time lag analysis illustrate that the different photon fluxes of different physical origins reach the peak at various times. The radio flux density peak has the longest time lag (about 5 days) compared to the highest X-ray energy band (30 - 100 keV) peak. For a lower X-ray energy band, the radio flux density has a shorter time lag (e.g., 11 - 30 keV of 4 days, and 6- 10 keV of $\sim 1$ day). The time lag gradually becomes indistinct for the X-ray light curves below 10 keV. 

The X-ray flux in different bands would be distinguished by the different physical origins: disk emission, reflection component, and corona emission. The corona flux variation shown in \ref{relflux} also illustrates a peak in LHS, thus, we also compared the radio curve with the corona flux curve. The red line of Fig. \ref{ccf} shows that the time lag of radio flux compared to the corona flux is about three days. The reflection fraction $F_{fr}$ has the evolution, also presents a peak pattern at MJD 58518, and the radio time lag from the $F_{fr}$ curve is about two days. Thus, the evolution curves of different components reach the peak in the time order: the highest energy X-ray photons from the corona emission (30 - 100 keV), relatively lower energy X-ray photons from the corona emission (11 - 30 keV), the total corona flux (including the photons reflected by the disk), and a reflection component fraction, then $r_{in}$ reaching ISCO about 1 day before the radio peak.

\section{discussion}\label{dis}

\begin{figure}
    \centering
    \includegraphics[scale=0.5]{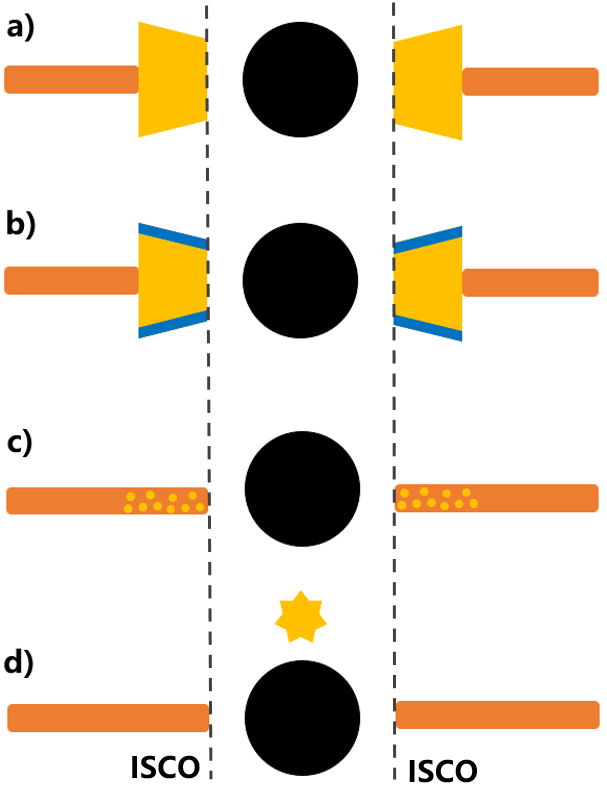}
    \caption{These cartoons show the disk-corona coupling and variations in the state transitions from LHS, HIS to SIS. The black circle represents the black hole, the yellow part represents the corona (the blue part around the corona is the cooled surface of the corona), the orange part is the accretion disk and dashed black line is ISCO. (a) In LHS, the ADAF/LHAF corona (the yellow part) exists within the inner edge $r_{in}$ of the accretion disk (the orange part). (b) The surface of the corona cooled down so that the cooled surface shaded more flux of the corona to the observer directly than the incident flux on the disk from the corona during HIS. (c) At the beginning of SIS (around MJD 58519) the disk is extending to ISCO and the cooling hot accretion flows would transit into two phases of the cold dense disk embedding hot gas (yellow circles). (d) During SIS the disk extends to ISCO, the reflection could be explained by the lamppost geometry.}
    \label{cartoon}
\end{figure}

The evolution of the reflection fraction and corona emission suggested a transition in the geometry of the corona during the LHS and HIS based on the Insight-HXMT observations. The corona near the BH \citep{yuan2014hot} could be advection-dominated accretion flows (ADAF, inside the accretion disk) or suitable for the lamppost model. The wide band X-ray spectral analysis finds that in the early stage of LHS, the inner edge of the accretion disk $r_{in}> 5$ ISCO, and after MJD 58518, $r_{in}$ is near ISCO. Therefore, ADAF that occupied the region within the inner edge of the accretion disk in the early stage, which is shown in Fig. \ref{cartoon} (a), disappeared after MJD 58518 with disk component flux of $\sim 1.6\times 10^{-8} erg/cm^{2}/s$ and corona flux of $5.2\times 10^{-8} erg/cm^{2}/s$. If we assume the black hole mass of 8 $M_{\odot}$ \citep{jana2020accretion} and the distance of this source is 2.2 kpc \citep{chauhan2021measuring}, then when $r_{in}$ reaches ISCO with the total emission of 10\% of Eddington luminosity, the disk component contributes to 2\% of Eddington luminosity, and the corona component contributes to 8\% of Eddington luminosity. 

In LHS, ADAF could also change to luminous hot accretion flows (LHAF), due to the increase of the accretion rate. For LHAF, despite the strong cooling process, compressional heating (the energy advection plays this role) can keep the accretion flow hot, which also requires a higher accretion rate. The maximum accretion rate of ADAF and LHAF are $\frac{\dot{M}_{crit,ADAF}}{\dot{M}_{Edd}}\sim 0.4\alpha^{2}$ and $\frac{\dot{M}_{crit,LHAF}}{\dot{M}_{Edd}}\sim \alpha^{2}$ respectively, where $\dot{M}_{crit}$ means the maximum accretion rate of this kind of accretion, $\dot{M}_{Edd}$ for the Eddington accretion rate, $\alpha$ for the dimensionless viscosity coefficient parameter\citep{narayan1995advection,nakamura1996global,yuan2001luminous}. This source has strong disk winds in LHS with a velocity $\sim 10^{4}\rm km/s$ \citep{wu2022accretion}, and despite $\alpha\sim$ 0.1 in general assumption, the simulation works \citep{igumenshchev2000two} suggest that $\alpha$ could reach and exceed 0.3 when there is a strong disk wind. In this source, the high-velocity wind comes from the inner region of the disk, which allows a higher $\alpha\sim 0.3$, then the critical accretion rate of LHAF and ADAF is $\sim$ 10\%$\dot{M}_{Edd}$ and $\sim$ 4\%$\dot{M}_{Edd}$, respectively. The luminosity range of the corona component in LHS and HIS increases from 3\% Eddington luminosity to 10\% Eddington luminosity, which exceeds the limitation of ADAF, and close to the limitation of LHAF. And when $r_{in}$ reaches ISCO (corona still has about 8\% Eddington luminosity), the high accretion rate of the thin disk would destroy LHAF.


When the thin disk reaches ISCO, there is still strong reflection from the disk, which implies the corona will not be located along the plane of the accretion disk. Thus, the corona could be explained by the lamppost model above the black hole. $F_{fr}$ increases from MJD 58517 and reaches 2 in MJD 58518 , while the incident flux decreases ($N_{re}$ represents the incident flux on the disk from the corona), therefore $F_{fr}$ rises due to the flux directly from corona dropping rapidly more than the flux incident on the disk, which implies that the obscuration of the corona by the material leads to the rapidly dropping off of the flux direct on the observer. The surface of the corona cooled down so that the cooled surface shaded more flux of the corona to the observer directly than the incident flux on the disk from the corona as shown in Fig. \ref{cartoon}(b) about one day before the thin disk occupying ISCO. In addition, the contribution of the reflected spectral component is higher than that of the direct emission, except in the highest energy band (see Fig. \ref{ref}). 

The radio flux increased and reached the peak about one day after the thin disk reached ISCO, whereas, the jets always company with the hot accretion flows. The hot accretion flow begins to cool and vanish probably as following physical picture\citep{yuan2003luminous}: the cooling process of hot accretion flows would lead to the transition into two phases of the cold dense disk embedding hot gas in Fig. \ref{cartoon} (c). The cold dense disk provides enough optical depth to reflect, and the hot gas as the ADAF continues to supply the material to jets through the jet-ADAF model \citep{yuan2002jet}. The different temperature parts of the corona may cool at different speeds: the higher temperature part corresponds to the faster cooling. Thus, there exist different time lags between X-ray and radio curves as illustrated in Fig. \ref{ccf}. Furthermore, the radio flux density peak lags the time of $r_{in}$ getting ISCO over one day, demonstrating that the two-phase cooling process lasts for about one day. After MJD 58520, the radio flux decreased which means that the jet weakens, however, the reflection still exists and the disk extended to ISCO, the corona could be explained by the lamppost geometry as shown in Fig. \ref{cartoon} (d).

Comparing the multiwavelength X-ray and radio flux evolution, with radio flux increasing to the peak, the accretion rate of the BH is around 10\% Eddington accretion rate or higher, which would help to understand radio origin and jet production mechanism. Magnetically arrested disk (MAD), as a kind of hot accretion flow that can exist within the accretion disk, is proposed to possibly produce radio emission in the BH systems \citep{xie2019radiative}. However, the inner radius of the accretion disk reaches near ISCO when the radio flux density is still high, thus the region of the hot accretion has disappeared, which suggests that the radio flux cannot come from the MAD. In addition, MAD also requires the maximum accretion rate at $\sim$ 3\% Eddington accretion rate \citep{xie2019radiative}, which is lower than the Insight-HXMT observations during LHS and HIS. The radio emission should have the origin of jets \citep{blandford1979relativistic}.

In this source during outburst, the observations may not support the BP jet mechanism \citep{blandford1982hydromagnetic,komissarov2007magnetic} and the BZ jet mechanism \citep{penna2013general,blandford1977electromagnetic,komissarov2007meissner} in MAX J13480-630. The BP jet mechanism produces the jet by extracting energy from the accretion disk, which needs the magnetic field through the accretion disk. The BZ process needs a high magnetic flux threading the horizon to extract the energy from the fast-rotating BH (however, only a moderate spin BH located in MAX J13480-630, \citealt{wu2023moderate}). The favorable situation of both mechanisms needs the magnetic flux of the disk near the limitation of MAD $\sim 50\rm G\ cm^{-2}$ corresponding to a magnetic field strength $\sim 10^{10}\ G$, which would lead the magnetic pressure to dominate the gas pressure and radiation pressure when the magnetic field would make thin disk thicker and the dissipation advect to the BH\citep{skadowski2016thin}. Our scenario prefers the magnetic tower mechanism \citep{lynden2003discs,hawley2006magnetically} in which the differential rotation accretion flows produce the high magnetic pressure gradient driving the gas accelerated away from the surface of the accretion disk to form the jet. The magnetic tower forms at a relatively high height and the high magnetic pressure gradient behaves in the normal direction of the accretion disk so that this mechanism has a small influence on the thin disk, allowing the existence of a relatively high accretion rate. 

\section{summary}\label{con}

In this paper, we report the spectral analysis of MAXI J1348-630 during the LHS and HIS of the 2019 outburst from 2 keV to 100 keV based on Insight-HXMT data. We have tested the spectrum with the power law and thermal Compton models, and there exist the obvious board iron line and Compton hump, which imply a strong reflection. The reflection model can well describe the spectra during the entire period. The inner edge of the accretion disk $r_{in}$ exceeded 5 ISCO during the LHS, and $r_{in}$ reaches at ISCO at MJD 58518 when the source was in HIS. The corona radiation showed a strong evolution during the LHS and HIS. The corona temperature evolved from $kT_{e}\sim 120\rm keV$ to $\sim 40\rm keV$ in LHS, and after MJD 58517 $kT_{e}$ returned to $\sim 100\rm keV$. During most periods of LHS and HIS, the reflection fraction stayed below 1, after MJD 58517, the reflection fraction increased to $\sim 1$ and reached $\sim 2$ at MJD 58518. The disk emission continued to increase during the observation, while the corona emission rose at first and decreased in HIS, showing a peak around MJD 58516.

Radio emission shows the time lags compared to hard X-ray emissions and reflection features. The high energy X-ray flux (30 - 100 keV) arrives at peak earliest about 5 days ahead of radio flux, and X-ray flux of 11-30 keV about 4 days ahead of radio flux, the reflection fraction ($F_{fr}$) also show a peak about 2 days ahead of radio flux. In addition, the inner radius of the accretion disk reaches the ISCO about 1 day before the radio peak. These disk-corona-jet coupling and evolution provide the clues of understanding corona geometry and jet production process. The hot inner accretion flow along the plane evolves from ADAF to LHAF, transiting to the corona explained by the lamppost model when $R_{in}$ is near ISCO. With the shrinkage of the hot accretion flow, the accretion rate increases to 10\% Eddington rate and radio emission also increases, favoring the magnetic tower jet mechanism instead of the MAD formation, which requires a lower accretion rate than 3\%.

We also propose a scenario for the disk-corona-jet coupling during the state transitions from LHS, HIS to SIS for this source. In LHS, the ADAF/LHAF corona exists within the inner edge of the accretion disk. With the surface of the corona cooling (HIS), the cooled surface shaded more flux of the corona to the observer directly than the incident flux on the disk from the corona. Around the start of SIS, the disk extends to ISCO, the jet is still powerful, the cooling process of hot accretion flows would lead to the transition into two phases of the cold dense clumps embedding hot gas. During the SIS, the jet weakens, the disk is near ISCO but the reflection could be explained by the lamppost geometry. The present scenario based on the reflection models has some simple theoretical assumptions, thus the more realistic model is expected in future.

\section*{Acknowledgements}
We are grateful to the referee for the useful comments and suggestions to improve the manuscript. This work is supported by the National Key Research and Development Program of China (Grants No. 2021YFA0718503 and 2023YFA1607901), the NSFC (12133007). This work has made use of data from the \textit{Insight-}HXMT mission, a project funded by the China National Space Administration (CNSA) and the Chinese Academy of Sciences (CAS).

\appendix
Observational information is listed in Table \ref{ob} including the observation ID, observation date (UTC \& MJD), the exposure time of three telescopes (LE, ME, HE), and the abbreviation of each observation.

\begin{table*}[h]
\centering
\caption{Insight-HXMT observation details for MAXI J1348-630. }
\label{ob}
\begin{tabular}{lcccccr} 
    \hline
    Obs. ID & Obs. date & Obs. date & LE Exposure(s) & ME Exposure(s) & HE Exposure(s) & Abbreviation   \\
	    & (yyyy-mm-dd) & (MJD) &  &  \\
    \hline
    P021400200101   & 2019-01-27 & 58510 & 2327.461 & 2327.461 & 2848.061 & 101\\
    P021400200201   & 2019-01-28 & 58511 & 1609.949 & 1609.949 & 2547.424 & 201\\
    P021400200301   & 2019-01-29 & 58512 & 2652.155 & 2652.155 & 3668.126 & 301\\
    P021400200401   & 2019-01-30 & 58513 & 2567.898 & 2567.898 & 3454.549 & 401\\
    P021400200403   & 2019-01-30 & 58513 & 2281.115 & 2281.115 & 3115.648 & 403\\
    P021400200404   & 2019-01-30 & 58513 & 1942.241 & 1942.241 & 2787.582 & 404\\
    P021400200405   & 2019-01-30 & 58513 & 1753.85  & 1753.85  & 2481.715 & 405\\
    P021400200407   & 2019-01-31 & 58514 & 1312.408 & 1312.408 & 797.5142 & 407\\
    P021400200408   & 2019-01-31 & 58514 & 1933.67  & 1933.67  & 2466.319 & 408\\
    P021400200410   & 2019-01-31 & 58514 & 1644.985 & 1644.985 & 2044.544 & 410\\
    P021400200411   & 2019-01-31 & 58514 & 2023.728 & 2023.728 & 2830.101 & 411\\
    P021400200412   & 2019-01-31 & 58514 & 1981.533 & 1981.533 & 2797.051 & 412\\
    P021400200415   & 2019-02-01 & 58515 & 1767.28  & 1767.28  & 2664.21  & 415\\
    P021400200416   & 2019-02-01 & 58515 & 1462.672 & 1462.672 & 1538.375 & 416\\
    P021400200417   & 2019-02-01 & 58515 & 1066.87  & 1066.87  & 847.3768 & 417\\
    P021400200418   & 2019-02-01 & 58515 & 2131.231 & 2131.231 & 3102.183 & 418\\
    P021400200419   & 2019-02-01 & 58515 & 2043.387 & 2043.387 & 2918.853 & 419\\
    P021400200420   & 2019-02-01 & 58515 & 1485.255 & 1485.255 & 2103.471 & 420\\
    P021400200501   & 2019-02-02 & 58516 & 1598.218 & 1598.218 & 2308.631 & 501\\
    P021400200601   & 2019-02-04 & 58518 & 1364.654 & 1364.654 & 378.9088 & 601\\
    P021400200602   & 2019-02-04 & 58518 & 2162.398 & 2162.398 & 2790.599 & 602\\
    P021400200603   & 2019-02-04 & 58518 & 1010.76  & 1010.76  & 1106.21  & 603\\
    P021400200701   & 2019-02-05 & 58519 & 239.40   & 1608.515 & 101.413  & 701\\
    P021400200702   & 2019-02-05 & 58519 & 325.18   & 2140.116 & 2910.508 & 702\\
    P021400200804   & 2019-02-06 & 58520 & 624.43   & 1911.50  & 2117.33  & 804\\
    P021400200904   & 2019-02-07 & 58521 & 239.40   & 1332.88  & 839.89   & 904\\
    P021400201101   & 2019-02-08 & 58522 & 638.4    & 2454.94  & 3210.02  & 1101\\
    \hline
\end{tabular}
\end{table*}

\bibliography{sample631}{}
\bibliographystyle{aasjournal}



\end{document}